# A Novel Method For Obtaining A Better Quality Speech Signal For Cochlear Implants Using Kalman With DRNL And SSB Technique

Rohini S. Hallikar[1], Uttara Kumari[2] and K Padmaraju[3]

[1] Department of ECE, R.V. College of Engineering, Bangalore, India
rohini_hallikar@yahoo.co.in
[2] Department of ECE, R.V. College of Engineering, Bangalore, India
[3] JNT University, Kakinada, Andhra Pradesh, India

*ABSTRACT*

*Cochlear implant devices are known to exist since a long time. The purpose of the present work is to develop a speech algorithm for obtaining robust speech. In this paper, the technique of cochlear implant is first introduced, followed by discussions of some of the existing techniques available for obtaining speech. The next section introduces a new technique for obtaining robust speech. The key feature of this technique lies in the use of the advantages of an integrated approach involving the use of an estimation technique such as a kalman filter with non linear filter bank strategy, using Dual Resonance Non Linear(DRNL) and Single Side Band(SSB) Encoding method. A comparative study of the proposed method with the existing method indicates that the proposed method performs well compared to the existing method.*

## 1. INTRODUCTION

Cochlear Implants importance lies in the fact that, these devices provide partial hearing for those who are profoundly deaf. Specialized techniques are required for speech processing and this involves the understanding of many disciplines such as signal processing, bioengineering, otolaryngology, physiology, speech science etc. One important objective of these signal processing techniques is, to derive electrical stimuli from the speech signal.[1]

For cochlear implants to work satisfactorily, robust auditory nerve fibres is an important requirement. Auditory nerve fibres, once they are stimulated, they impart the associated neural impulses to the brain, which is interpreted as sound by the brain. The loudness of a sound is directly proportional to two important factors, rates of firing and number of nerve fibres that are activated. The implant conveys sound information to the brain. This information also conveys the loudness of the sound which further depends upon amplitude of the stimulus current and pitch of the sound. [2]

## 2. SIGNAL PROCESSING TECHNIQUES

User's hearing potential is maximized by the use of cochlear implant speech coding techniques. Techniques such as Multipeak, spectral maxima, Spectral peak, continuous interleaved sampling (CIS) etc are the examples of existing techniques. Above mentioned techniques do facilitate the cochlear implant users, to obtain their hearing and then communicating in quiet. But these techniques are at a disadvantage when complex listening has to take place. [3]

In complex listening, the speech coding technique has to achieve a balance with respect to the quality of the hearing and reduction in the noise level. This arrangement effectively reduces the effect of noise and at the same time gives better signal restoration. Also the use of a nonlinear filter bank model gives a robust formant representation and speech perception in noise. A Dual Resonance non Linear (DRNL) model is comparatively simpler than other adaptive non-linear models of the basilar membrane [4]. To achieve these objectives a kalman filter based on the voice generation model and a DRNL with SSB filter are used.

## 3. PROPOSED TECHNIQUES

The proposed method is shown in figure 1. Input from the microphone is fed to an estimation algorithm such as a kalman filter. The processed signal is given to a preemphasis stage. From the preemphasis stage the signal is fed to a DRNL filter block.
Output of DRNL filter is given to a SSB encoder followed by an envelope detection stage to obtain the biphasic pulses. This is followed by the reconstruction of the input signal. The input waveform is then compared with the reconstructed one on the basis of the correlation coefficient. The above processing is done initially considering clean speech and later corrupting the signal with noise.

The following representation gives the step by step representation of the proposed method.
Step 1 Input from the microphone to be given to kalman filter.
Step 2 The output of first step to be preemphasized.
Step 3 The preemphasized signal is given to a DRNL filter block.
Step 4 Output of the DRNL filter to be given to an SSB encoder
Step 5 SSB encoded signal to be given to an envelope detector.
Step 6 Get biphasic pulses from the envelope detector. These biphasic pulses are used to
obtain the reconstructed signal.

## 4. RESULTS

For various inputs of speech, correlation coefficients of the proposed method along with existing method were computed. The correlation coefficient between the input signal and reconstructed waveform was computed by using (1).

$$r = \frac{\sum_i (x_i - x')(y_i - y')}{[\sum_i (x_i - x')^2 (y_i - y')^2]^{0.5}} \quad (1)$$

Where, r is the correlation coefficient between two symbols x and y, with x' and y' representing their means.

Table 1. Comparison of correlation coefficient for two different methods Fs=10KHz for the speech signal 'news.wav'

| Listening conditions | Proposed Method (1) | DRNL(2) |
|---|---|---|
| Quiet | 0.8026 | 0.7888 |
| 5 dB babble noise | 0.4713 | 0.4377 |
| 10 dB babble noise | 0.5475 | 0.4768 |

Table 2. Comparison of correlation coefficient for two different methods Fs=20 KHz for the speech signal 'news.wav'

| Listening conditions | Proposed Method (1) | DRNL(2) |
|---|---|---|
| Quiet | 0.7940 | 0.7658 |
| 5 dB babble noise | 0.4609 | 0.4136 |
| 10 dB babble noise | 0.5375 | 0.4562 |

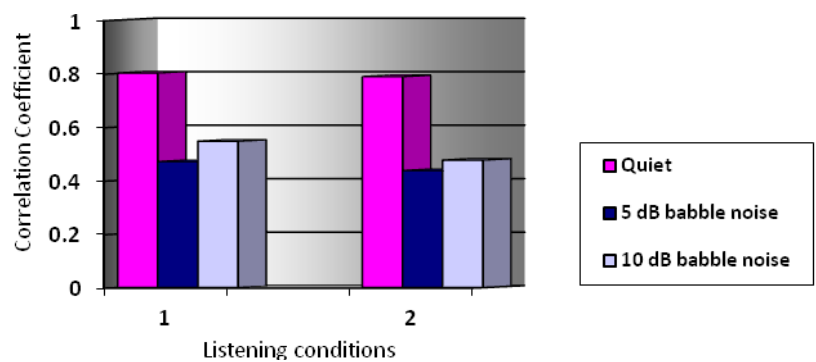

Fig 2: The performance of proposed method & DRNL for different listening conditions for news.wav using Fs=**10 KHz**

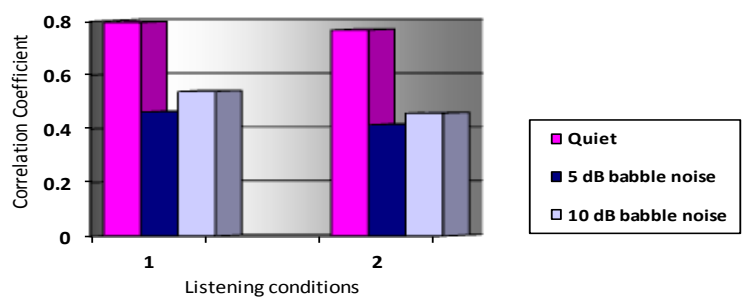

Fig 3: The performance of proposed method with and DRNL methods for different listening conditions for news.wav using Fs=**20 KHz**

Table 3: Comparative analysis of the improvement of proposed method over existing method for the speech signal 'news.wav' Fs=**10KHz**

| Listening conditions | Improvement of correlation coefficient of Proposed method over existing method(%) |
|---|---|
| Quiet | 1.749 |
| 5 dB babble noise | 7.6 |
| 10 dB babble noise | 14.828 |

Table 4: Comparative analysis of the improvement of proposed method over existing method for the speech signal 'news.wav' Fs=**20 KHz**

| Listening conditions | Improvement of correlation coefficient of Proposed method over existing method(%) |
|---|---|
| Quiet | 3.682 |
| 5 dB babble noise | 11.43 |
| 10 dB babble noise | 17.82 |

Table 3 and Table 4 convey the improvement of the proposed method over existing method. The proposed method showed good improvement over existing method for quiet as well as noisy conditions.

## 5. CONCLUSION

The key contribution of this paper is the implementation of a new method which gives a superior performance in quiet and noisy conditions.

**Authors**

Rohini S.Hallikar
Completed  B.E. (Electronics) degree from
Dr.B.A.M.U Aurangabad, Maharashtra.
Completed M.Tech (Digital Electronics &
Communication), VTU, Belgaum.  Currently
working as Assistant Professor in department of
Electronics and Communication, R.V.College
of Engineering,
 Bangalore. 560059, .Karnataka, India

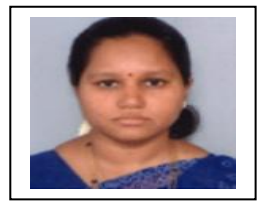

Dr..M.Uttara Kumari
Completed  B.Tech(E&C) from Nagarjuna
University. Completed M.E.(Signal Processing,
Communication) from Bangalore University.
Completed PhD  from Andra University.
Presently, working as the  Head of the
Department  in department of Electronics and
Communication,R.V.College of  Engineering,
 Bangalore. 560059, .Karnataka, India

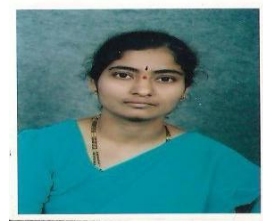

Dr. K. Padma raju
Completed B.Tech (E & C)  from Nagarjuna
University.. Completed M.Tech(Electronic
Instrumentation), National Institute of
Technology Warangal. Completed PhD from
Andra University. Completed Post
Doc.Fellowship from Hoseo University South
Korea.
      Currently Principal and Professor of
Electronics and Communication Engineering
and   Jawaharlal Nehru Technological
University Kakinada .
Kakinada - 533 003, Andhra Pradesh, INDIA

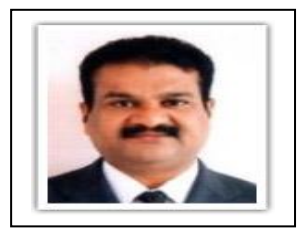